\documentclass[twoside]{article}
\usepackage{amsmath}
\usepackage[silent,pdf,hyper]{simpletexml0}
\usepackage{hyperref}
\usepackage{epsfig}
\usepackage{comment}

 \journalid{ES}{15}{4}{15}
\citnumber{}{1}
 \paperid{2015ES000558}
\cpright{}{2015}
\lefthead{reshetnyak}
\righthead{Inverse problem}
\siterooturl{http://rjes.wdcb.ru}
\xmldepositor{}{}    
\xmlregistrant{}     

\gcmdterms{}  
\aguterms{}  
   \keywords{mean-field dynamo;  magnetic field;   $\alpha$-, $\omega$-effects;  reversals}

\authaddr{M. Yu. Reshetnyak,
Schmidt Institute of Physics of the Earth of the Russian Academy of Sciences,
  Moscow, Russia, 
  
    Pushkov Institute of Terrestrial Magnetism, Ionosphere and Radio Wave Propagation of the Russian Academy of Sciences, Moscow, Russia, 
 \noindent (m.reshetnyak@gmail.com)}

\begin{document}

\title{Inverse problem in Parker's dynamo}

   \author{1}{Reshetnyak}{M. Yu.}{}{1}{2}{}

\affil{1}{Schmidt Institute of Physics of the Earth of the Russian Academy of Sciences}{Moscow}{}{Russia}
\affil{2}{Pushkov Institute of Terrestrial Magnetism, Ionosphere and Radio Wave Propagation of the Russian Academy of Sciences}{Moscow}{}{Russia}

 \abstract{ The inverse solution of the 1D Parker dynamo equations is considered. The method is based on minimization of the cost-function, which characterize deviation of the model solution properties from the desired ones. The output is the latitude distribution of the magnetic field generation sources: the $\alpha$- and $\omega$-effects. Minimization is made using the Monte-Carlo method. The  details of the method, as well as some applications, which can be interesting for the broad dynamo community, are considered: conditions when the invisible for the observer  at the surface of the planet toroidal part of the magnetic field is much larger than the poloidal counterpart. It is shown  that at some  particular distributions of  $\alpha$ and $\omega$  the well-known thesis that  sign of the dynamo-number defines equatorial symmetry  of the magnetic field to the equator plane, is violated. It is also  demonstrated in what circumstances magnetic field in the both hemispheres  have different properties, and simple physical explanation of this phenomenon is proposed.      \vskip 0.1cm}

 \section{1. Introduction}  

The observed magnetic field in the various astrophysical objects, like planets, stars and galaxies,  is a product of the dynamo mechanism. The dynamo theory,  which first success was concerned with the development of the mean-field dynamo [{\it  Krause and R\"{a}dler}, \reflink{KR80}{1980}], to the present time transformed to the new branch of   physics, and combined recent  knowledges on the structure and evolution of the objects, fluid dynamics, supercomputer modeling. To now it  can describe many typical features of the magnetic field, known from  observations [{\it  R\"udiger et al.}, \reflink{RKH2013}{2013}], [{\it   Roberts and King}, \reflink{RK2013}{2013}].

As it usually happens during the  development of the new theory, the  first approach is the direct solution of the model  equations with prescribed parameters, which are chosen due to some a priori information on the system. Whether it  leads to  the acceptable  correspondence of the model with the  observations,  the fine tuning of the model parameters starts. This is the subject of the inverse problem, where basing on the observations, and usually on the fixed equations, the governing parameters of the model are looked for.

There are different ways how it can be done. Here we consider approach, where the desired parameters are the forms of the spatial distribution of the energy sources in the dynamo equations. We limit our study to the  simple, but well-adopted in the dynamo community, 1D Parker's equations with the algebraic quenching, which are traditionally used in the planetary, galactic, and stellar  dynamo applications [{\it  R\"udiger et al.},  \reflink{RKH2013}{2013}]. These equations describe  evolution of the axi-symmetric   mean magnetic field, which depends on the latitude $\vartheta$. The sources of the energy, the $\alpha$- and $\omega$-effects, are the prescribed functions of $\vartheta$. The aim is to find such  distributions of $\alpha$ and $\omega$ in $\vartheta$, which satisfy some restrictions on the simulated magnetic field. The measure of deviation of the model from the desired state is characterized by the cost-function $\Psi$. To minimize numerical expenses  we decompose $\alpha$ and $\omega$ in the Fourier series  in the polar angle $\theta=\pi/2-\vartheta$, and rewrite $\Psi$ in terms of the spectral coefficients, where only the first $N$ modes are used.  Minimization of $\Psi$, which can have  quite complex structure,  should be done using some robust method. So far $\Psi$ usually has local  minima, we used modification of the Monte-Carlo method, the good candidate for the parallel simulations at  the cluster supercomputer systems,  used in the work.

\vspace*{.7ex}
Below we consider some examples, which demonstrate implementation of the method, and show how information on the spatial spectrum of the magnetic field, its periodicity, ratio of the poloidal and toroidal magnetic energies can be used for the estimates of the optimal profiles of $\alpha$ and $\omega$. We stress attention that the inverse approach in dynamo applications is very rare, compared to the direct simulations, and only a few papers in this direction exist.

  \section{2. Dynamo in the  sp herical shell} 

We consider simple dynamo model in the spherical shell
  [{\it  Ruzmaikin et al.}, \reflink{RSS1988}{1988}]:

\begin{equation}\label{1}
\begin{array} {l}\displaystyle
     {\partial A\over  \partial  t} =\alpha B
    + \widehat{L}\, A  \\ \\ \displaystyle
      {\partial B\over  \partial  t} =-{\Omega} {\partial \over\partial\theta} A+ \widehat{L}\,B,
\end{array}
\end{equation}

\noindent   where   $A$ and $B$ are  the azimuthal components of the vector potential $\bf A$,
  and magnetic field ${\bf B}={\rm rot}\,{\bf A}$, $\alpha(\theta)$ is the $\alpha$-effect;
   ${\Omega}(\theta)$ is the differential rotation, and
     $\displaystyle \widehat{L}=\eta\left({1\over \sin\theta}{\partial \over\partial\theta} \sin\theta {\partial \over\partial\theta} -{1\over\sin^2\theta}\right)$ is the diffusion operator with $\eta$ for the magnetic diffusion.
  System \eqref{1} is solved in the interval $0\le \theta\le \pi$ with the boundary conditions
  $B=0$ and $A=0$ at $\theta=0$ and $\pi$.

To exclude the exponentially growing solution of Eqs(1) 
the $\alpha$-quenching is used.
    The form of quenching depends on the particular objects. In planetary and galactic dynamos the simple algebraic form is acceptable. In  the solar dynamo the dynamical quenching is usually used, see for details
    [{\it  Kleeorin et al.}, \reflink{KRR}{1995}].

       Here we consider the local form of the algebraic $\alpha$-quen\-ching:
\begin{equation}\label{2}
\begin{array}{l}\displaystyle
   \displaystyle \alpha(\theta)={\alpha_\circ(\theta)\over 1+E_m},
 \end{array}
\end{equation}

\noindent with the
            magnetic energy  $E_m(\theta)=(B_r^2+B^2)/2$, and radial component of the magnetic field
  $\displaystyle B_r={1\over \sin\theta}{\partial\over\partial\theta}\left(\sin\theta\, A\right)$.

 \section{3. Inverse problem}    

The direct solution of the system \eqref{1},\eqref{2} with the prescribed profiles of $\alpha_\circ(\theta)$ and $\Omega(\theta)$  gives  $\bf B(\theta,\, t)$, which can be compared with the observations. The disadvantage  of the direct problem  is a pure knowledge on  $\alpha_\circ(\theta)$ and $\Omega(\theta)$. Thus, in the planetary dynamo these profiles are known only from 3D simulations, see, e.g.,
      [{\it  Reshetnyak}, \reflink{R10}{2010}]. For the solar dynamo  [{\it  Belvedere et al.},\reflink{Bel00}{2000}] information on $\Omega$ comes from the helioseismology, however $\alpha$-effect is still varies from model to model. In galactic dynamo situation is similar to the solar dynamo, that is why the simplest models of $\alpha_\circ$ are still so popular.
   These reasons motivate the inverse problem approach, where different profiles of $\alpha_\circ(\theta)$ and $\Omega(\theta)$ are tested on  observations. 

       Let introduce the cost-function $\Psi({\bf B},\, {\bf B^o})$, where $\bf B$ is the model magnetic field, and $\bf B^o$ is the observable one. Then $\Psi$  has at least one minimum at ${\bf B}={\bf B^o}$.
     The proper choice of $\Psi$, and sufficient observations $\bf B^o$ make this minimum global.
      Usually, observations do not cover the whole  domain of the magnetic field generation, either one observes such  properties of the magnetic field that magnetic field can not be recovered in the unique way. Then  $\Psi$ has local minima as well, and for minimization of $\Psi$ one requires special efficient methods, see review in
 [{\it  Press et al.}, \reflink{NR2007}{2007}].

    The next step is to simplify the problem and consider only the large-scale features of profiles, e.g., the first $N_\alpha,\,N_\Omega$ Fourier harmonics in $\theta$:
\begin{eqnarray*}
 \begin{array}{l}\displaystyle
              \displaystyle \alpha_\circ=\sum\limits_{n=1}^{N_\alpha} C_n^\alpha\sin(2\theta n),\qquad
        \displaystyle \Omega=\sum\limits_{n=0}^{N_\Omega} C_n^\Omega\cos(2\theta n).
 \end{array}
\end{eqnarray*}

     Then, the  problem reduces to the search of  such ${\bf C^\alpha}$ and ${\bf C^\Omega}$ that $\Psi({\bf C^\alpha},\,{\bf C^\Omega})$ has minimum (maybe local). In general case, study of  the sequence of minima, obtained during simulations, is  interesting too.

       The numerical  details of the direct solver, based on the central $2^{nd}$-order finite-difference approximation of the spatial derivatives, and $4^{th}$-order Runge-Kutta method for integration in time, are described in
             [{\it  Reshetnyak}, \reflink{R14}{2014}]. The direct  C++ solver was wrapped, using MPI interface, so that at each computer node the direct problem \eqref{1},\eqref{2} for the different  $({\bf C^\alpha},\, {\bf C^\Omega})$, given by the random generator, was solved.

            The random Gauss generator,  with the mean value,  equal to the previous best choice, and
     standard deviation  $3\sigma$,
       generates set of $({\bf C^\alpha},\, {\bf C^\Omega})$. It is supposed that  $({\bf C^\alpha},\, {\bf C^\Omega})$  should be in the fixed region.
          After selection of  $({\bf C^\alpha},\, {\bf C^\Omega})$
             at the current iteration step, which corresponds to the minimal $\Psi$, the new
              $({\bf C^\alpha},\, {\bf C^\Omega})$ were generated, and then the next iteration started.
           The shift of the mean value of
                 $({\bf C^\alpha},\, {\bf C^\Omega})$, which is optional, helps to  increase convergence of the process. This method is  modification of the Monte-Carlo method, see  the  basic ideas in   [{\it  Press et al.}, \reflink{NR2007}{2007}].

                To solve equations
                at
                  Lomonosov's supercomputer in Mos\-cow State University and the Joint Supercomputer Center of RAS,
                   $N=101$ grid points for the spatial approximation, the time step $\tau=10^{-5}$, and  ${\cal N}$  computer nodes from 10 to 100 for parallelization
                were used.  Usually, number of  iterations was less than 10, and depended on $\cal N$. Application of MPI and cluster computers
                for 1D problem is not crucial, but it will be of great importance for the 2D code (with radial dependence), which is under  development now.


   Further we consider some particular forms of $\Psi$ and discuss the resulted profiles of $\alpha_\circ(\theta)$ and $\Omega(\theta)$ in details.

\section{4. Ratio of the Pol oidal and Toroidal Magnetic Energies}

The measure of intensity of generation sources in \eqref{1} is the so-called dynamo-number, defined as: $\displaystyle {\cal D}={||\alpha_\circ|| \, ||\Omega||\, L^3\over \eta^2}$, where $L=\pi$ is the spatial scale, and $||.||$ is the  norm of the physical quantity, discussed below.
                    Here we consider how solution of  \eqref{1},\eqref{2}, with fixed  $||\alpha_\circ||$ and $||\Omega||$, depends on the forms of  profiles.

\begin{figure}[ht !]   
 \def \ss {9cm}   
   \figurewidth{20pc}
    \vskip -0.0cm
 \hskip -0.25cm \epsfig{figure=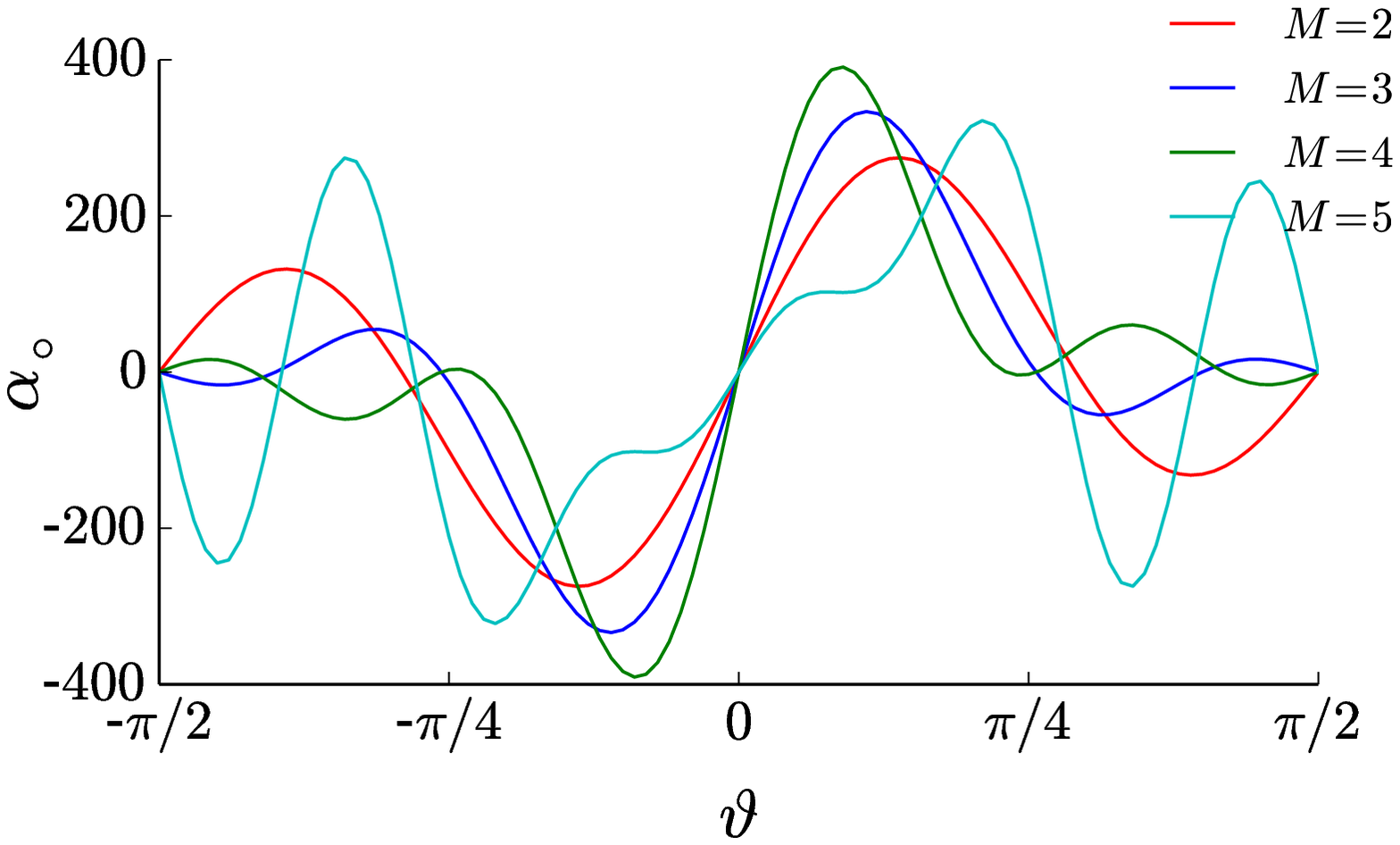,width=\ss}
 \vskip -0.0 cm
 \hskip -0.25cm \epsfig{figure=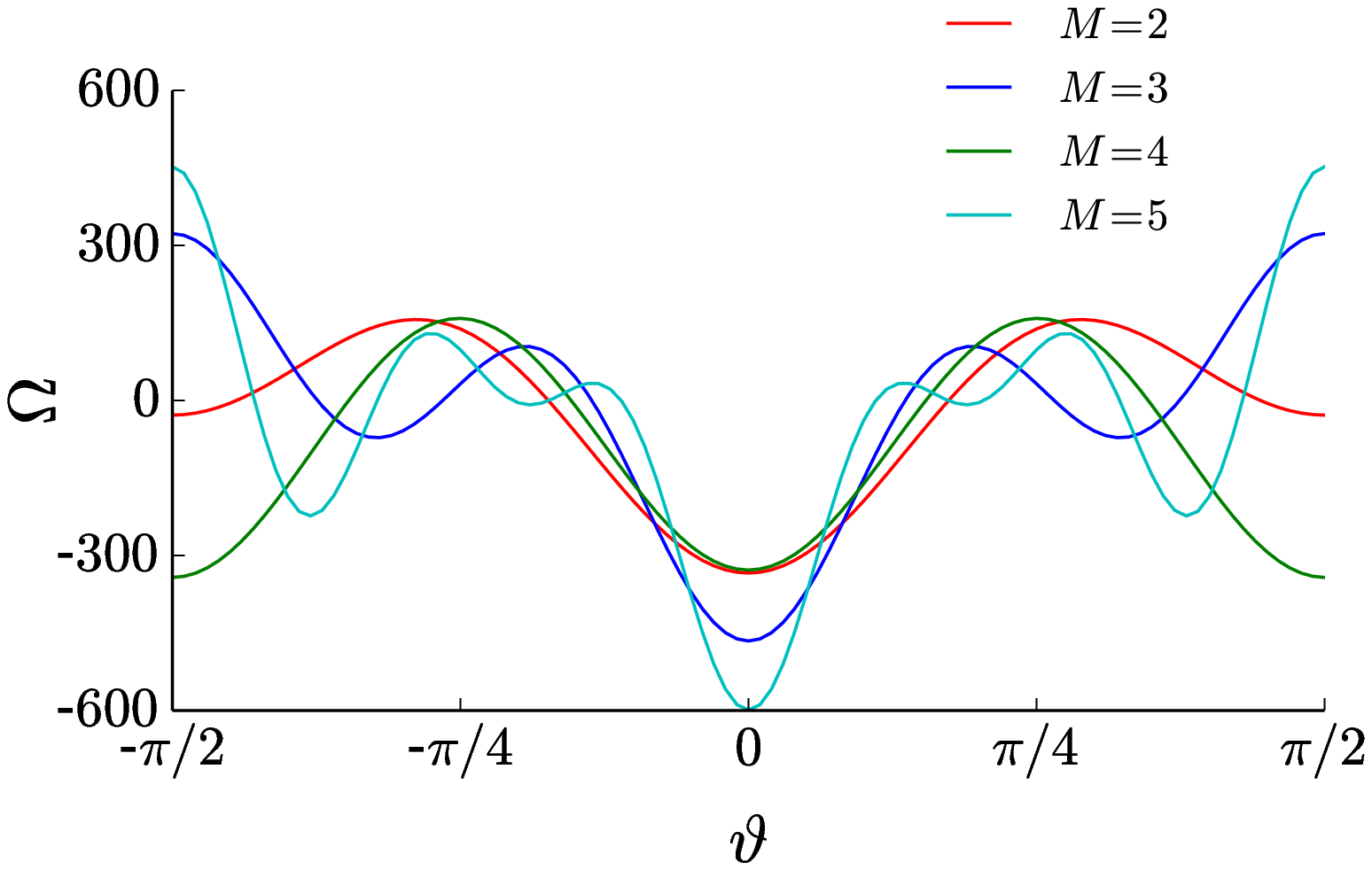,width=\ss}
  \vskip -0.0 cm
  \hskip -0.25cm \epsfig{figure=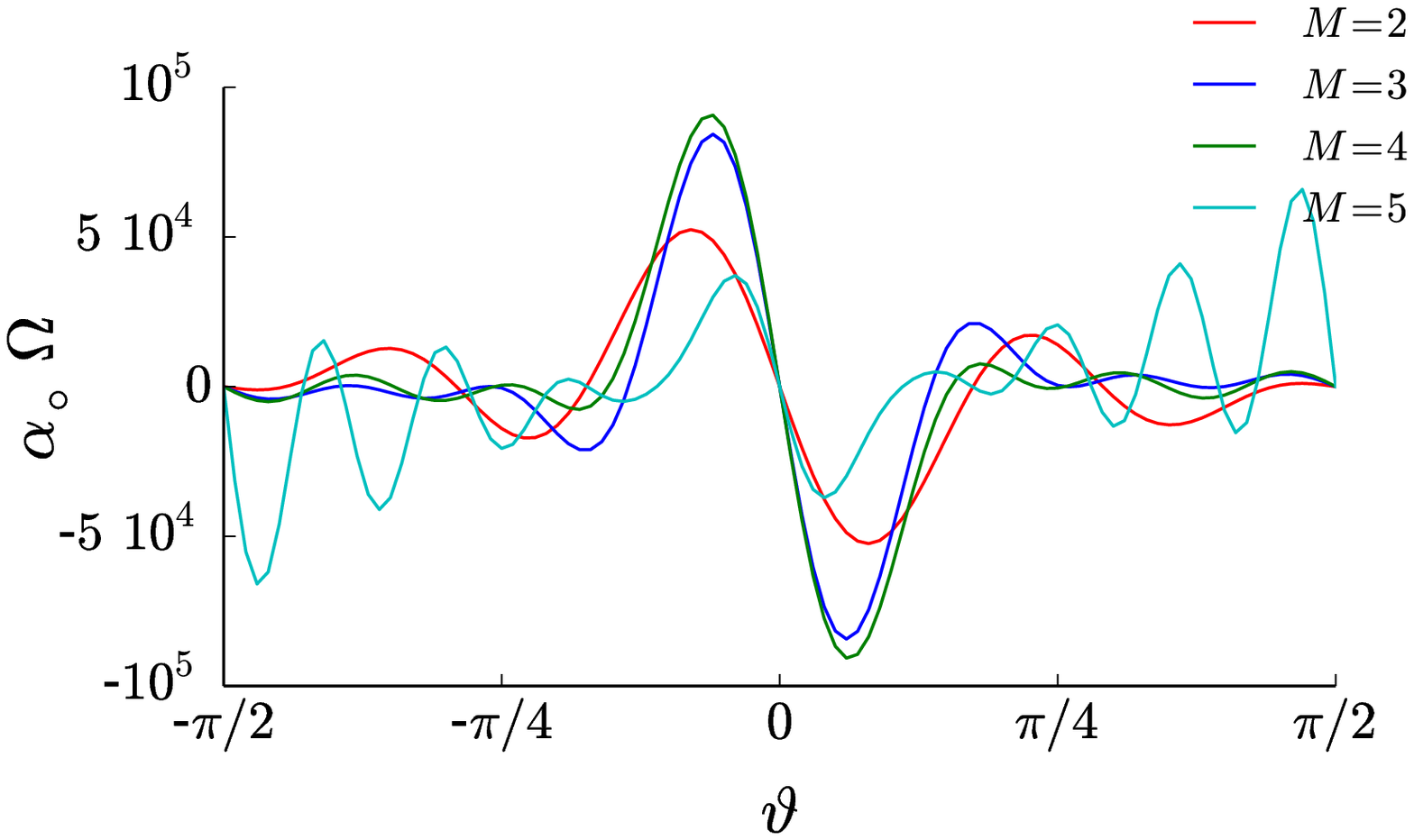,width=\ss}
  \vskip -0.3cm
  \caption{
   Latitude dependence of $\alpha_\circ$, $\Omega,$ and their product $\alpha_\circ\Omega$ for minimal $\Psi$.  } \label{1}
 \end{figure}
   Having in mind that the both quantities $\alpha_\circ$, $\Omega$, can change the sign, we introduce the following definitions of norms: \\
 $||{\alpha}_\circ||=\pi^{-1}\int\limits_o^\pi |\alpha_\circ|\sin\theta\, d\theta$,  and $||\Omega||  =\pi^{-1}\int\limits_o^\pi |\Omega|\sin\theta\, d\theta $.

\begin{figure}[ht!]   
 \def \ss {9cm}   
   \figurewidth{20pc}
    \vskip -0.0cm
 \hskip -0.25cm \epsfig{figure=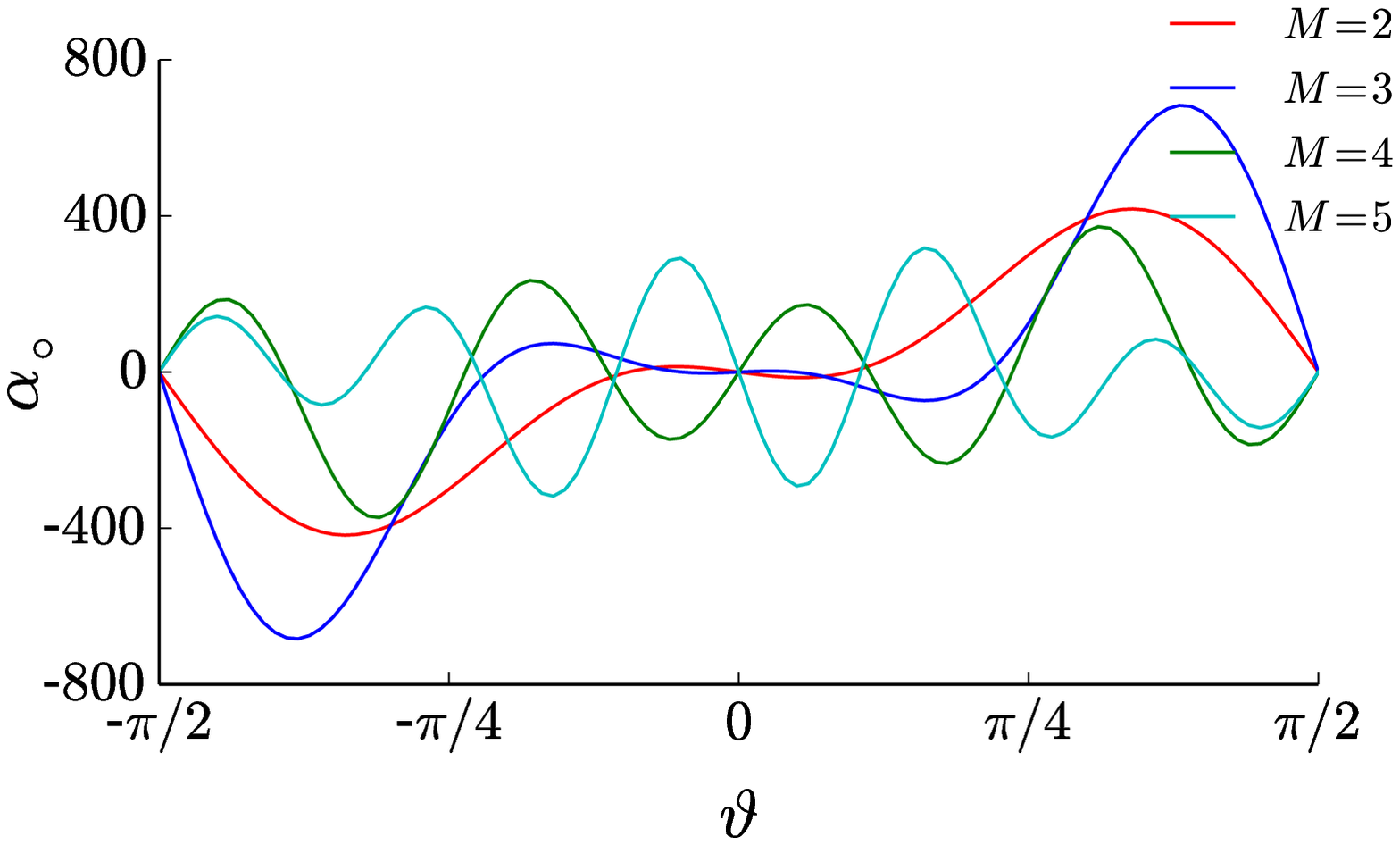,width=\ss}
 \vskip -0.0 cm
 \hskip -0.25cm \epsfig{figure=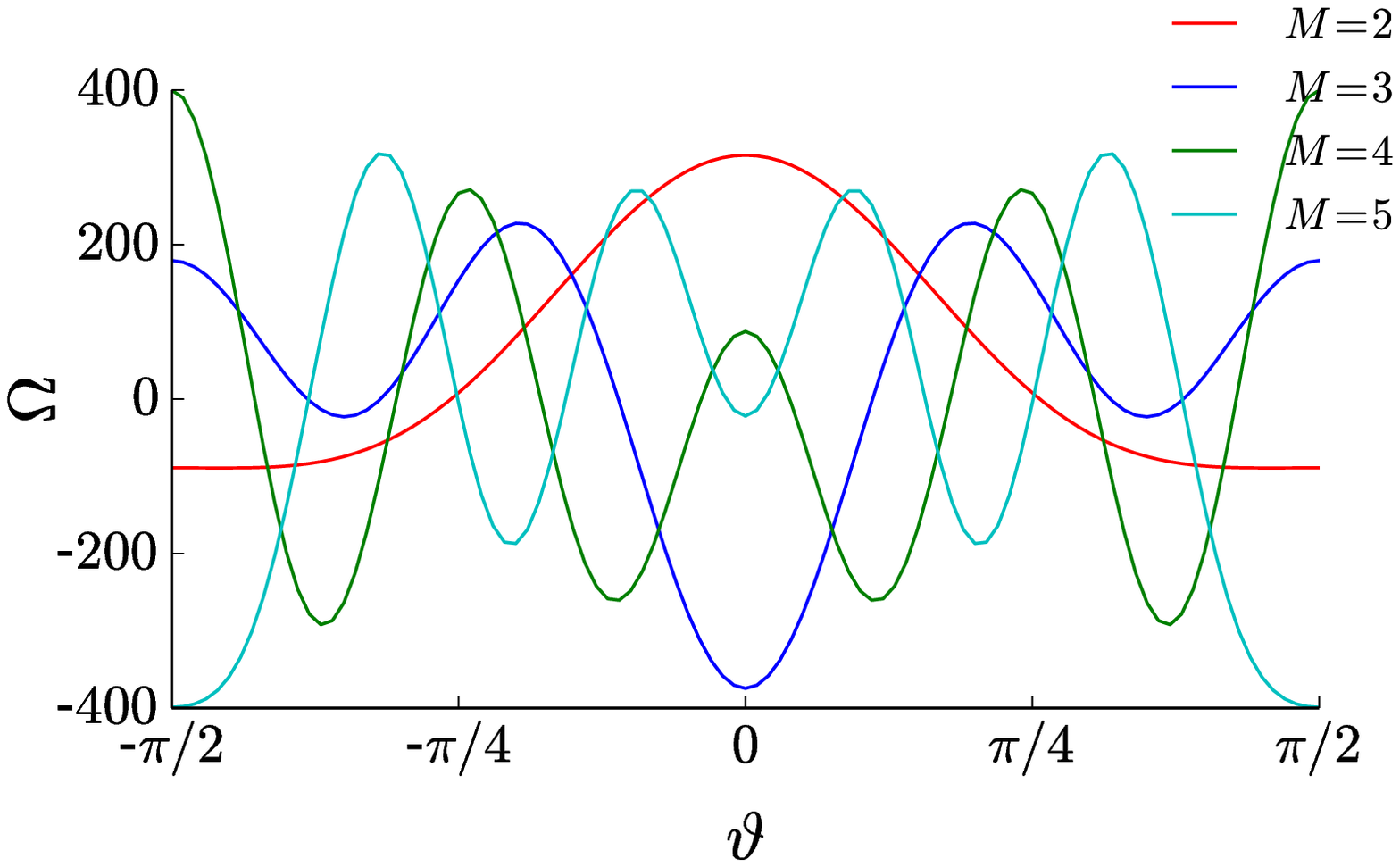,width=\ss}
  \vskip -0.0 cm
  \hskip -0.25cm \epsfig{figure=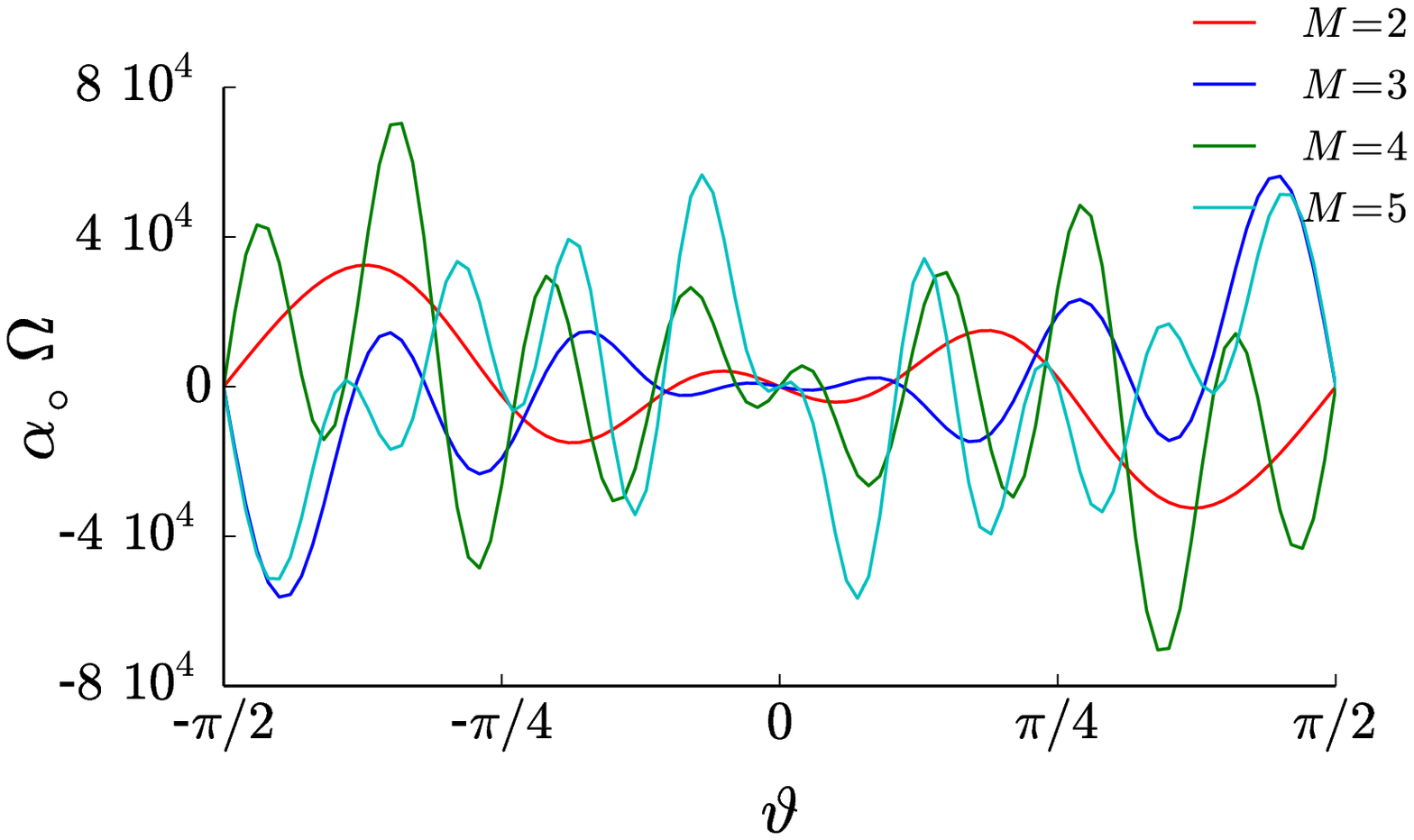,width=\ss}
  \vskip -0.3cm
  \caption{
  Latitude dependence of $\alpha_\circ$, $\Omega,$ and their product $\alpha_\circ\Omega$ for maximal $\Psi$.  } \label{2}
 \end{figure}

We look for  such solutions of \eqref{1}, \eqref{2}, which for the fixed  $||\alpha_\circ||$, $||\Omega||$, have minimal, either maximal ratio $\cal R$ of
 the poloidal $\displaystyle E_p^T=B^2/2$   and toroidal   $\displaystyle E_p^T=B^2/2$  magnetic energies.
   So far in many astrophysical applications only the one component of the  magnetic field  (poloidal or toroidal) can be observed, estimate of the whole magnetic energy $\displaystyle E_m=E_m^T(1+{\cal R})$ can vary from model to model, and amplitude of its variations is  the subject of active debates
[{\it  Brandenburg and Subramanian}, \reflink{BS05}{2005}].

   Simple analysis of \eqref{1},\eqref{2} leads to the following predictions:\vspace*{.5\baselineskip}
\begin{equation}\label{3}
\displaystyle
       {  \cal R} =
  \begin{cases} \displaystyle
    {||\alpha_\circ||\over||\Omega|| L},    & \quad E_m \ll 1\\ \\ \displaystyle
 {\eta^2 \over ||\Omega||^2 L^4},
    & \quad E_m\gg 1,
  \end{cases}
\end{equation}\vspace*{.5\baselineskip}

\noindent that follows to that $\cal R$ is defined by $||\alpha_\circ||$ and  $||\Omega||$.
       Our aim is to find dependence of $\cal R$ on these profiles.

  Let  introduce the cost-function
    $\displaystyle \Psi=1-e^{-{\cal R}}$, and find $({\bf C}^\alpha,\, {\bf C}^\Omega)$, which extremum of $\Psi$.
 Latitude distributions of  $\alpha_\circ$ and $\Omega$ for  the  four cases $N_\alpha=N_\Omega=M$ with $M=2\dots 5$, and
  $||\alpha_\circ||=10^2$, $||\Omega||=10^2$, are  presented in
  \figref{1} and \figref{2}.

             Firstly note that some details  do depend on $M$. This is natural for the small $M$. However usage of the large $M$ would be inconsistent with the basics of the mean-field dynamo, where the large-scale fields are considered. In other words, the number of harmonics $M$ should be much  less than the number of the grid points $N$ in the  numerical scheme for Eqs\eqref{1}. It means  that here we discuss  only  the large-scale trends  in the model, and they do exist.

              Before to start the  analysis of \figref{1} and \figref{2}, note that minimal (${\cal R}^{min}\sim 10^{-4}$) and maximal ${\cal R}^{max}\sim 1$ correspond to the different levels of the total magnetic energy $E_m$: for  ${\cal R}^{min}$ one has $E_m\sim 10^3$, and for ${\cal R}^{max}$ -- $E_m\sim 1$.
 In agreement with  estimate \eqref{3}, the case   ${\cal R}^{min}$  corresponds to $E_m\gg 1$.
   On contrary, in the case   ${\cal R}^{max}$,  by some reasons, there is  suppression of the total magnetic field generation.

                     Following further note that due to our normalization,  amplitudes  of $\alpha_\circ$, and $\Omega$, see \figref{1} and \figref{2},  do not demonstrate significant differences. But as was already mentioned before, the measure of the field generation is the product  $\cal D$. And this quantity does demonstrate the different behaviour for two branches.
                  For  ${\cal R}^{min}$ (large $E_m$) there is only one extremum of $\alpha_\circ\Omega$ in the hemisphere. This  helps to generate  the large-scale magnetic field.

                        For  ${\cal R}^{max}$ (small $E_m$) the product $\cal D$ oscillates in $\theta$ coordinate. The scale of the fields is smaller than in the  case of ${\cal R}^{min}$, and as a result, the magnetic diffusion is larger. Whether  for  ${\cal R}^{min}$ for all $M$, the leading harmonic for $B_r$ is stable quadrupole (Legendre polynomial with $l=2$), then for  ${\cal R}^{max}$ during the time  solution switches  from $l=1$ (dipole) to higher orders: even to  $l=10$ at $M=5$. So far the amplitudes of $\alpha_\circ$ and $\Omega$ are of the same order in the both cases, difference in $\cal R$ is a product of low correlation in space of
                       $\alpha_\circ$ and $\Omega$, as well as of the energy sources with the generated magnetic field. The first  option is  shown in
  \figref{1}, where the maximum of the product near the equatorial plane  is clearly pronounced. On contrary, this correlation is small in
  \figref{2}.
  It supports suggestion that  localizations of the  both energy sources ($\alpha_\circ$ and $\Omega$) in the same place helps to the large-scale magnetic field generation.

          The test on the field configurations reveals that for ${\cal R}^{min}$ the both components of the quadrupole magnetic field have maximum at the equator, so that in that region  the products of the magnetic field components and  $\alpha_\circ$,  $\Omega$ are large, and as a result, the magnetic field generation is  enhanced.

         For the case ${\cal R}^{max}$ correlation between the magnetic field and energy sources is weak, and efficiency of the dynamo mechanism is small.
   Situation can  change if the meridional circulation, providing transfer  of the magnetic field from one region of generation to the other, will be taken into account. Then effective generation of the magnetic field with the different localizations of $\alpha_\circ$ and $\Omega$ is possible.

\section{5. Pure Dipole and Non-dipole Solutions}

The another prediction of the linear analysis of Parker equations with simple forms of  $\alpha_\circ$ and $\Omega$ is that  alternation  of sign  $\cal D$ leads to the change of the symmetry of the leading mode: the dipole mode switches to the quadrupole, and vice versa. This  change can also be accompanied with transition  from the
     stationary to oscillatory regimes. Using our approach we test whether this prediction is valid for complex forms of $\alpha_\circ$ and $\Omega$ in the non-linear  regime.

   Let  introduce the  cost-function
      $\displaystyle \Psi=1-e^{-{g_1^2 / \sum\limits_{{l=2}}^{11}g_l^2}}$, where $g_l$ are  the spectral coefficients in decomposition on the Legendre polynomials. The  same norms   $||\alpha_\circ||$ and $||\Omega||$, as in the previous section,  were used.
             Minimum of $\Psi$ corresponds to the non-dipole configuration, and maximum limits to the pure dipole field, respectively. As we will see, the two groups with dipole ($l=1$) and non-dipole ($l>1$) configurations will dominate.

      The four runs with $M=2\dots 5$ for minimal and maximal $\Psi$ were done. For maximal $\Psi$ the stationary dipole solution was observed for  all the runs. The toroidal energy was $E_m\approx 650$, and the poloidal one was   two orders less. Exception was the case with $M=5$ with  $E_m^T\approx 100$, and ${\cal R}\approx 0.1$.

   The regimes with minimal $\Psi$ demonstrated various behaviour in time. Cases with $M=2,\, 4$ were  the stationary quadrupoles with ${\cal R}\approx 0.1$
         and $0.01$, and $E_m\approx 900$, $40$, respectively. In the case  $M=3$ we got   ${\cal R}\approx 1$, $E_m\approx 10$. The dominant oscillatory mode was $g_l=6$. The last stationary regime with $M=5$ corresponded to $g_l=4$.

         The visual analysis of product $\alpha_\circ\,\Omega$ does not reveal any significant differences between the  branches of the minimal and maximal $\Psi$. To test whether the sign $\cal D$ plays the role, we calculated integrals $\int\limits_0^{\pi/2} \alpha_\circ\Omega\, d\theta$, for  $\Psi^{min}$:
           $4.3\, 10^4$, $2.7\, 10^4$, -$4\, 10^3$, -$2\, 10^4$, and for  $\Psi^{max}$:   $3.4\, 10^4$, $3.2\, 10^4$, $2.1\, 10^4$, -$5\, 10^3$. As we can see,
          the sign of the integral does not influence on whether solution  is  dipole, either it is  quadrupole. Moreover, there is no correlation of sign of  $\cal D$ with the symmetry of the magnetic field  over the equator plane in the non-linear regime. This result demonstrates once more how predictions of the linear analysis should  be used carefully in the saturated states.

\section{6. Dynamo-wave Through  Equator}

The  asymmetry of the magnetic fields over the equator plane is  well-known to observers.  In geomagnetism this problem was discussed in  [{\it  Gubbins et al.}, \reflink{GBGL2000}{2000}], where the idea of  the
 interplay of the dipole and quadrupole modes was proposed. These two modes have similar thresholds of generation and its
superposition  can enforce the total magnetic field in one hemisphere, and  weaken it in the other one.
 The paleomagnetic records, often based on the assumption of the axial dipole, can not exclude   this possibility even for  Phanerozoic.

In the solar dynamo asymmetry presents at least in two forms: the difference between the magnetic fluxes from two hemispheres is finite, and can change the sign in time
 [{\it  Knaack et al.}, \reflink{KSB2004}{2004}]. The other remarkable phenomenon is that during the Maunder minimum in the $17^{th}$ century more than 95\% of the   sunspots were located in the southern hemisphere of the Sun
 [{\it  Ribes and Nesme-Ribes}, \reflink{RN1993}{1993}].

 Another  example of the break of the magnetic field  equatorial symmetry demonstrates Mars's crustal field
  [{\it  Stanley et al.}, \reflink{SEZP2008}{2008}]. This field is associated with the internal magnetic field generated by the dynamo mechanism in the past.

      The equatorial asymmetry of the magnetic field is allowed by the dynamo theory as well. The 3D dynamo simulations can reproduce  this phenomenon for the particular set of parameters
 as for the spherically symmetrical  boundary conditions
 [{\it  Grote and Busse}, \reflink{GB2000}{2000}],  [{\it  Busse and Simitev}, \reflink{FH2006}{2006}],
 [{\it  Landeau and Aubert}, \reflink{LA2011}{2011}],
 as well as for the
 heterogeneous  heat flux at the outer boundary of the spherical shell
  [{\it  Stanley et al.}, \reflink{SEZP2008}{2008}],
  [{\it  Amit et al.}, \reflink{ACL2011}{2011}],
[{\it Dietrich and Wicht}, \reflink{DW2013}{2013}].

It should be noted that  possibility of such asymmetries is also interesting from the general point of view. It motivates us to use the
  inverse approach to   test this phenomenon at the simple dynamo-model.

           In assumption that dynamo wave, say for the field $A$, is monochromatic, its phase velocity is   $\displaystyle V_A={A'_t/ A'_\theta}$. Information on $V_A$ can be used to distinguish between the two cases:  the  wave, which propagates through the equator plane, either it  vanishes at the plane, and then recovers with the opposite  sign in the second hemisphere\footnote{The third possibility, when the wave is reflected from the equator plane, is not supported by the observations.}.

 The mean value of $V_A$ in the equatorial band $\vartheta=\pm \vartheta_b$ is
   $\left< V_A \right> = \int\limits_{-\vartheta_b}^{\vartheta_b} V_A\, d\vartheta$.
 In assumption, that the band is narrow enough, so that  $V_A$  changes (if does) the  sign only at the equator,
   the normalized quantity $\displaystyle {\cal F}= \left< V_A \right> / \left< \Big{|}V_A\Big{|} \right>$ ranges in the interval $[0,\, 1]$.
 The case  ${\cal F}=0$ corresponds to the vanishing wave at the  plane  $\vartheta=0$. The second extreme case is  $|{\cal F}|=1$, when  $V_A$ has the same
sign over the whole band.

\begin{figure}[ht!]  
  \def \ss {13cm}   
   \figurewidth{20pc}
       \vskip -5.2cm
 \hskip -0.7cm \epsfig{figure=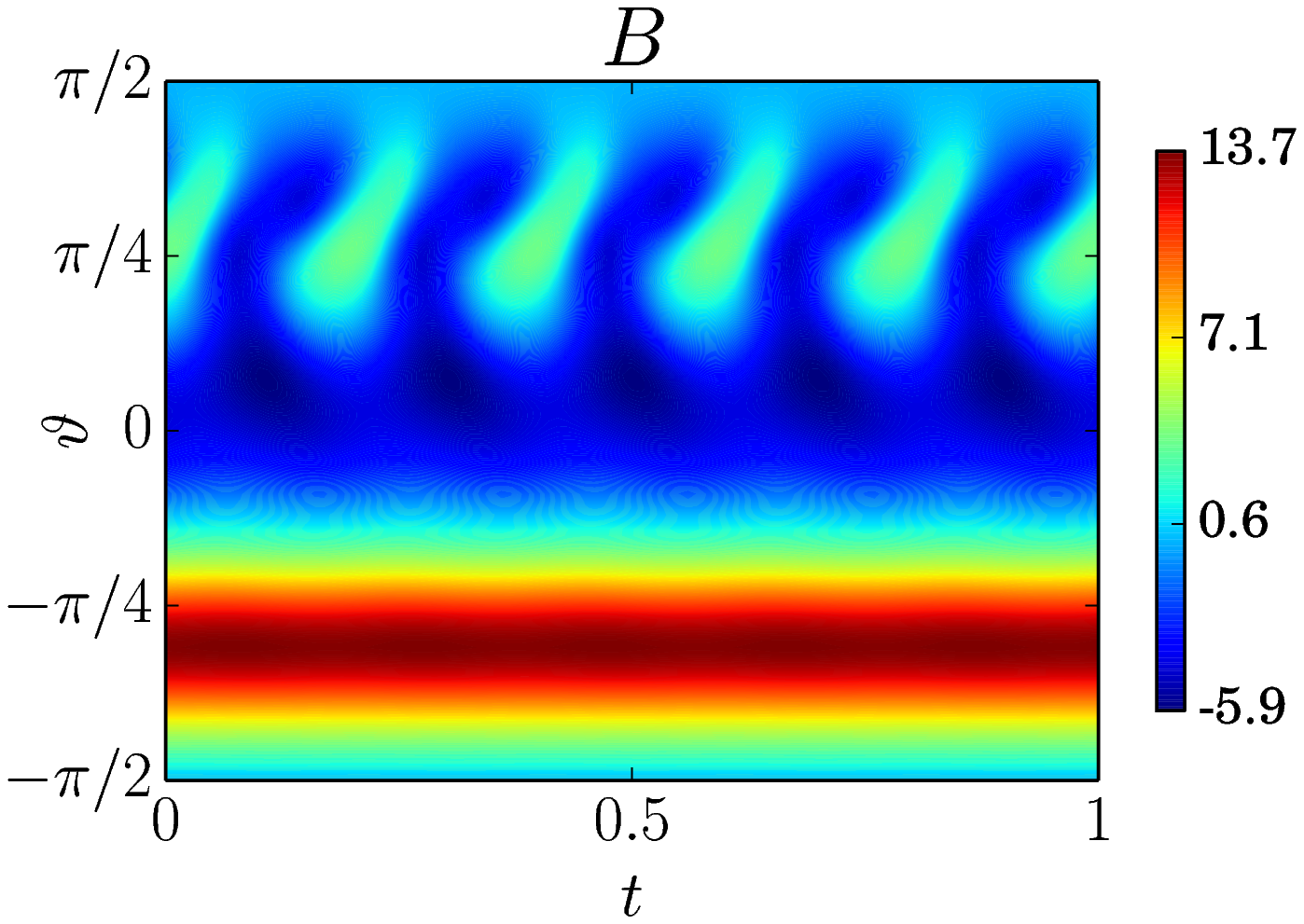,width=\ss}
  \vskip -10.5 cm
  \hskip -0.7cm \epsfig{figure=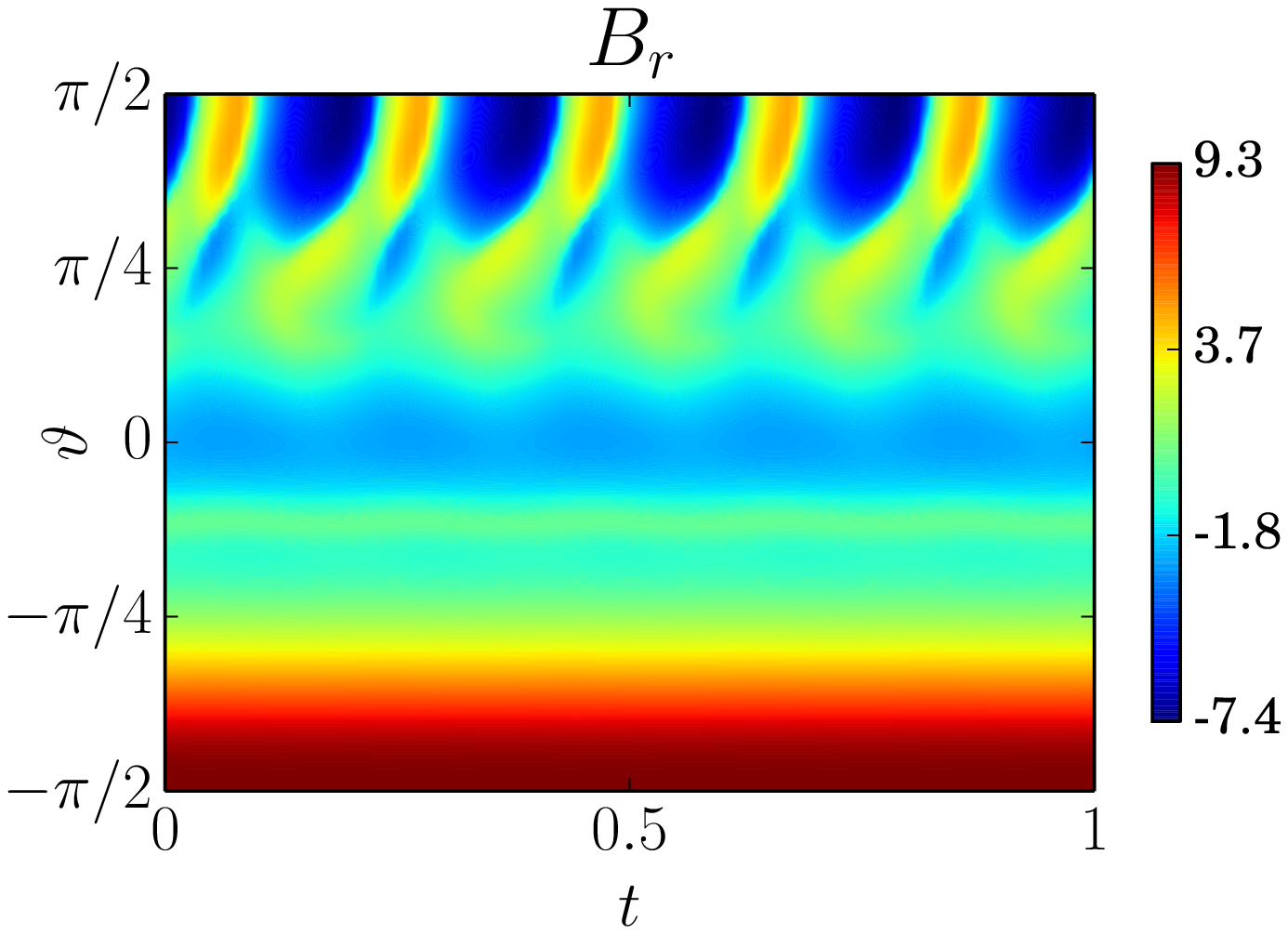,width=\ss}
     \vskip -10.5 cm
     \hskip -0.7cm \epsfig{figure=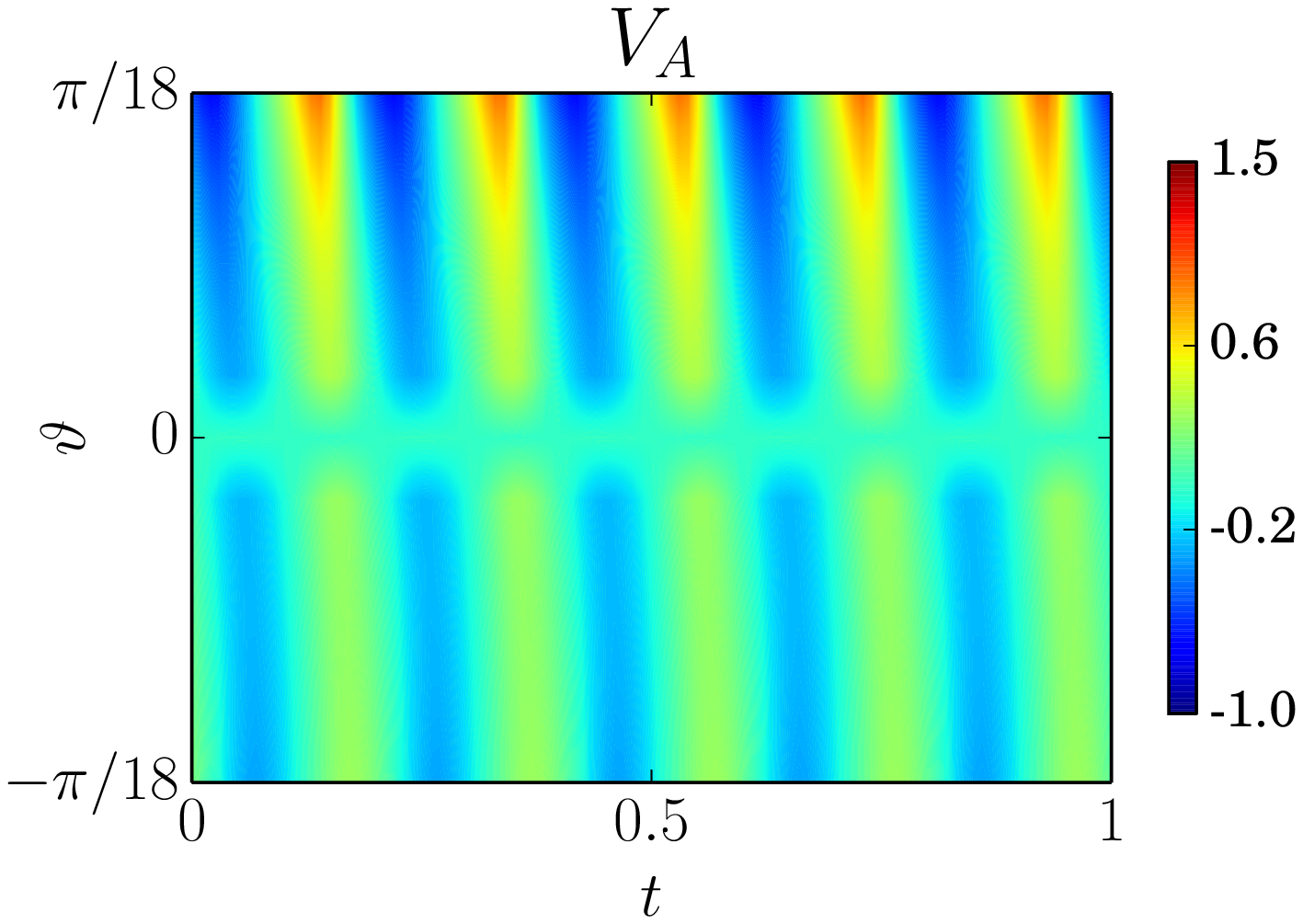,width=\ss}
   \vskip -5.5cm
    \caption{The butterfly diagrams for $B$, $B_r$ components of the magnetic field, and phase velocity $V_A$ of the poloidal  magnetic field for $M=2$ and $||\alpha_\circ||=|\Omega||=50$.} \label{3}
 \end{figure}

The proposed cost-function has the following form:
\begin{equation}\label{4}
    \displaystyle
     \Psi ={1\over 2}\left(e^{-|{\cal F}|} + e^{-{\cal G}}
     \right).
\end{equation}

The first term in the sum in \eqref{4} corresponds to the described above restriction on the wave behaviour   in the band. The second term helps to filter out the non-oscillatory solutions:

\begin{eqnarray*}
    \displaystyle
      {\cal G}={f_1 \over f_1+f_2}, \qquad f_1=\overline{E_m-\overline{E_m}},\,\qquad f_2=\overline{E_m},
\end{eqnarray*}

\noindent  where the overline means  averaging over the whole space and time. The case with ${\cal G}\ll 1$ corresponds to the small amplitude oscillations, compared to the  mean level of $E_m$. We do not interesting in this regime.
       The case with ${\cal G}\sim  1$ corresponds to the large oscillations: e.g., for $E_m=\sin^2(\nu \theta)$, and any integer
      $\nu$, ${\cal G}  \approx 0.68$.

 The largest $\cal F$ and $\cal G$ provide minimum of $\Psi$ in \eqref{4}.

  The simulated magnetic field, see the butterfly diagrams in \figref{3}, demonstrate the
         quite different behaviour in the northern and southern hemispheres. In the northern hemisphere it consists of two kinds of waves, which travel to the poles at the high latitudes, and from the poles to the equator in the band $\vartheta=\pm \pi/4$. There are periodic reversals of the magnetic field, which correspond to the change of the sign of $B_r$.

    On the contrary,        in the southern hemisphere the main part of the  magnetic field is constant in time. The poloidal field $B_r$ is concentrated near the pole, and maximum of the toroidal field $B$ is shifted to $\vartheta\approx-\pi/3$.

   This quite strange configuration of the magnetic field, at least compared to the usual field states, corresponds to the class of  the hemispherical dynamo, mentioned in the beginning of the section. Note that we did not use any imposed asymmetry in the model, and this result is the intrinsic property of the model, as it was discovered in some 3D simulations.

  Returning to the way how we get this solution, we remind that the crucial point was selection of regimes with the non-zero mean phase velocity $V_A$ of the radial magnetic field in the equatorial band, see
                  \figref{3}. There  are waves of $V_A$, traveling from the north pole to the southern, with the constant magnitude, except the region near the  equator plane, where $V_A$   is small. If  resolution of observations  is pure, then it seems that the dynamo wave penetrates free  through the equator plane from one hemisphere  to the other. The direction of this wave changes in time, however the mean value of $V_A$ over the time and space domain in  \figref{3} is not zero. It is this deviation from the zero value the cost-function \eqref{4} detected.

    The possible explanation of our hemispherical dynamo is  concerned with the spatial distribution of $\alpha_\circ$ and $\Omega$, obtained in the inverse model, \figref{4}.

\begin{figure}[ht!]   
       \def \ss {8cm}   
      \figurewidth{20pc}
    \vskip -0.0cm
         \hskip -0.2cm \epsfig{figure=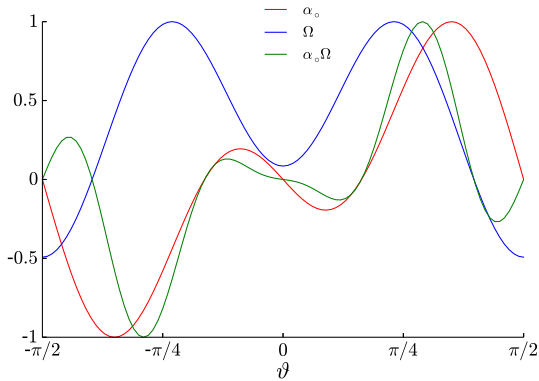,width=\ss}
     \vskip -0.5cm
  \caption{
      Latitude dependence of $\alpha_\circ$, $\Omega,$ and their product $\alpha_\circ\Omega$ with $M=2$.} \label{4}
 \end{figure}
    
       We observe coincidence of $\alpha_\circ$ and $\Omega$ extrema's locations. It results in the large product $\cal D$. Situation is similar to that one in  \figref{1}, where correlation of $\alpha_\circ$ and $\Omega$ was also strong. However, in that case extrema of $\cal D$ were near the equator plane, on contrary to  the hemispherical dynamo, where  they are shifted to the middle latitudes. It is this shift of maximum of the magnetic field generation helps to isolate dynamo process in hemispheres from each other, and permits  different evolutions of the magnetic field in the hemispheres. We emphasize that the observed flux, concerned with the phase velocity $V_A$, is quite small, and does not change situation substantially. But as we demonstrated, this flux is  the result of the equatorial symmetry break, which leads to the very different morphologies of the magnetic fields in the hemispheres.

 \section{7. Conclusions} 
\vspace*{-.5\baselineskip}

           Having deal with  the direct dynamo problem solutions, I  really enjoyed to work with  the inverse problem approach for  this toy dynamo model. In spite of the fact that 1D model itself is out of date, the level of abstraction in communication with the computer
           in  the inverse approach is much higher than in the direct problems. In the inverse approach
  one             formulates the properties of the desired solution, and then tries to understand why  the resulted parameters provide these properties. This process  is much more intriguing rather than to use the fixed parameters, and follow the  results of the direct problem, where  solution is already also prescribed.
  However the latter approach can be used for the more sophisticated models, its not the fact that the simpler model in inverse approach will not give  the  better result due to the finer tuning of parameters.

       The obtained above results are the product of numerous tries, when  for many times I wandered why the computer selected this or that particular regime. The lack of criteria, which were used for the cost-function construction, sometimes resulted in the very unexpected results. Many restrictions, which are supposed  by default, should be explained straightforward to the computer. However the results are worthy of these efforts. May be what is more important, is that this approach  stimulates understanding of the model. With  minimal number of criteria, it is possible to find scenarios, which can be tested, using more complex models. This inverse approach can be  useful tool for asking a good questions, even the answers would be quite wrong.
  As regards to the simplicity of the considered model, estimates of the required computer time shows that the inverse method, considered here,  can be extrapolated to the higher dimensional models as well.

\end{document}